\documentclass[%
superscriptaddress,
preprint,
 amsmath,amssymb,
 aps,
]{revtex4-1}

\usepackage{graphicx}
\usepackage{dcolumn}
\usepackage{bm}

\usepackage[mathlines]{lineno}

\newcommand{\del}{\partial}

\begin{document}


\title{Evolution of a plasma column measured through\\ modulation of a high-energy proton beam}

\author{S.~Gessner}
\affiliation{CERN, Geneva, Switzerland} 

\author{E.~Adli}
\affiliation{University of Oslo, Oslo, Norway}
\author{A.~Ahuja}
\affiliation{CERN, Geneva, Switzerland}
\author{O.~Apsimon}
\affiliation{University of Manchester, Manchester, UK}
\affiliation{Cockcroft Institute, Daresbury, UK}
\author{R.~Apsimon}
\affiliation{Lancaster University, Lancaster, UK}
\affiliation{Cockcroft Institute, Daresbury, UK}
\author{A.-M.~Bachmann}
\affiliation{CERN, Geneva, Switzerland}
\affiliation{Max Planck Institute for Physics, Munich, Germany}
\affiliation{Technical University Munich, Munich, Germany}
\author{F.~Batsch}
\affiliation{CERN, Geneva, Switzerland}
\affiliation{Max Planck Institute for Physics, Munich, Germany}
\affiliation{Technical University Munich, Munich, Germany}
\author{C.~Bracco}
\affiliation{CERN, Geneva, Switzerland}
\author{F.~Braunm{\"u}ller}
\affiliation{Max Planck Institute for Physics, Munich, Germany}
\author{S.~Burger}
\affiliation{CERN, Geneva, Switzerland}
\author{G.~Burt}
\affiliation{Lancaster University, Lancaster, UK}
\affiliation{Cockcroft Institute, Daresbury, UK}
\author{B.~Buttensch{\"o}n}
\affiliation{Max Planck Institute for Plasma Physics, Greifswald, Germany}
\author{A.~Caldwell}
\affiliation{Max Planck Institute for Physics, Munich, Germany}
\author{ J.~Chappell}
\affiliation{UCL, London, UK}
\author{E.~Chevallay}
\affiliation{CERN, Geneva, Switzerland}
\author{M.~Chung}
\affiliation{UNIST, Ulsan, Republic of Korea}
\author{D.~Cooke}
\affiliation{UCL, London, UK}
\author{H.~Damerau}
\affiliation{CERN, Geneva, Switzerland}
\author{G.~Demeter}
\affiliation{Wigner Research Center for Physics, Budapest, Hungary}
\author{L.H.~Deubner}
\affiliation{Philipps-Universit{\"a}t Marburg, Marburg, Germany}
\author{A.~Dexter}
\affiliation{Lancaster University, Lancaster, UK}
\affiliation{Cockcroft Institute, Daresbury, UK}
\author{S.~Doebert}
\affiliation{CERN, Geneva, Switzerland}
\author{J.~Farmer}
\affiliation{Heinrich-Heine-University of D{\"u}sseldorf, D{\"u}sseldorf, Germany}
\author{V.N.~Fedosseev}
\affiliation{CERN, Geneva, Switzerland}
\author{R.~Fiorito}
\affiliation{University of Liverpool, Liverpool, UK}
\affiliation{Cockcroft Institute, Daresbury, UK}
\author{R.A.~Fonseca}
\affiliation{ISCTE - Instituto Universit\'{e}ario de Lisboa, Portugal}
\author{F.~Friebel}
\affiliation{CERN, Geneva, Switzerland}
\author{L.~Garolfi}
\affiliation{CERN, Geneva, Switzerland}

\author{B.~Goddard}
\affiliation{CERN, Geneva, Switzerland}
\author{I.~Gorgisyan}
\affiliation{CERN, Geneva, Switzerland}
\author{A.A.~Gorn}
\affiliation{Budker Institute of Nuclear Physics SB RAS, Novosibirsk, Russia} 
\affiliation{Novosibirsk State University, Novosibirsk, Russia}
\author{E.~Granados}
\affiliation{CERN, Geneva, Switzerland}
\author{O.~Grulke}
\affiliation{Max Planck Institute for Plasma Physics, Greifswald, Germany}
\affiliation{Technical University of Denmark, Lyngby, Denmark}
\author{E.~Gschwendtner}
\affiliation{CERN, Geneva, Switzerland} 
\author{A.~Hartin}
\affiliation{UCL, London, UK}
\author{A.~Helm}
\affiliation{GoLP/Instituto de Plasmas e Fus\~{a}o Nuclear, Instituto Superior T\'{e}cnico, Universidade de Lisboa, Lisbon, Portugal}
\author{J.R.~Henderson}
\affiliation{Lancaster University, Lancaster, UK}
\affiliation{Cockcroft Institute, Daresbury, UK}
\author{M.~H{\"u}ther}
\affiliation{Max Planck Institute for Physics, Munich, Germany}
\author{M.~Ibison}
\affiliation{University of Liverpool, Liverpool, UK}
\affiliation{Cockcroft Institute, Daresbury, UK}
\author{S.~Jolly}
\affiliation{UCL, London, UK}
\author{F.~Keeble}
\affiliation{UCL, London, UK}
\author{M.D.~Kelisani}
\affiliation{CERN, Geneva, Switzerland}
\author{V.K.~Khudyakov}
\affiliation{Budker Institute of Nuclear Physics SB RAS, Novosibirsk, Russia}
\affiliation{Novosibirsk State University, Novosibirsk, Russia}
\author{S.-Y.~Kim}
\affiliation{UNIST, Ulsan, Republic of Korea}
\author{F.~Kraus}
\affiliation{Philipps-Universit{\"a}t Marburg, Marburg, Germany}
\author{M.~Krupa}
\affiliation{CERN, Geneva, Switzerland}
\author{T.~Lefevre}
\affiliation{CERN, Geneva, Switzerland}
\author{Y.~Li}
\affiliation{University of Manchester, Manchester, UK}
\affiliation{Cockcroft Institute, Daresbury, UK}
\author{S.~Liu}
\affiliation{TRIUMF, Vancouver, Canada}
\author{N.~Lopes}
\affiliation{GoLP/Instituto de Plasmas e Fus\~{a}o Nuclear, Instituto Superior T\'{e}cnico, Universidade de Lisboa, Lisbon, Portugal}
\author{K.V.~Lotov}
\affiliation{Budker Institute of Nuclear Physics SB RAS, Novosibirsk, Russia}
\affiliation{Novosibirsk State University, Novosibirsk, Russia}
\author{M.~Martyanov}
\affiliation{Max Planck Institute for Physics, Munich, Germany}
\author{S.~Mazzoni}
\affiliation{CERN, Geneva, Switzerland}
\author{V.A.~Minakov}
\affiliation{Budker Institute of Nuclear Physics SB RAS, Novosibirsk, Russia}
\affiliation{Novosibirsk State University, Novosibirsk, Russia}
\author{J.C.~Molendijk}
\affiliation{CERN, Geneva, Switzerland}
\author{J.T.~Moody}
\affiliation{Max Planck Institute for Physics, Munich, Germany}
\author{M.~Moreira}
\affiliation{GoLP/Instituto de Plasmas e Fus\~{a}o Nuclear, Instituto Superior T\'{e}cnico, Universidade de Lisboa, Lisbon, Portugal}
\affiliation{CERN, Geneva, Switzerland}
\author{H.~Panuganti}
\affiliation{CERN, Geneva, Switzerland}
\author{A.~Pardons}
\affiliation{CERN, Geneva, Switzerland}
\author{F.~Pe\~na~Asmus}
\affiliation{Max Planck Institute for Physics, Munich, Germany}
\affiliation{Technical University Munich, Munich, Germany}
\author{A.~Perera}
\affiliation{University of Liverpool, Liverpool, UK}
\affiliation{Cockcroft Institute, Daresbury, UK}
\author{A.~Petrenko}
\affiliation{CERN, Geneva, Switzerland}
\affiliation{Budker Institute of Nuclear Physics SB RAS, Novosibirsk, Russia}
\author{A.~Pukhov}
\affiliation{Heinrich-Heine-University of D{\"u}sseldorf, D{\"u}sseldorf, Germany}
\author{S.~Rey}
\affiliation{CERN, Geneva, Switzerland}
\author{H.~Ruhl}
\affiliation{Ludwig-Maximilians-Universit\"{a}t M\"{u}nchen, Munich, Germany}
\author{H.~Saberi}
\affiliation{CERN, Geneva, Switzerland}
\author{P.~Sherwood}
\affiliation{UCL, London, UK}
\author{L.O.~Silva}
\affiliation{GoLP/Instituto de Plasmas e Fus\~{a}o Nuclear, Instituto Superior T\'{e}cnico, Universidade de Lisboa, Lisbon, Portugal}
\author{A.P.~Sosedkin}
\affiliation{Budker Institute of Nuclear Physics SB RAS, Novosibirsk, Russia} 
\affiliation{Novosibirsk State University, Novosibirsk, Russia}
\author{A.~Sublet}
\affiliation{CERN, Geneva, Switzerland}
\author{P.V.~Tuev}
\affiliation{Budker Institute of Nuclear Physics SB RAS, Novosibirsk, Russia}
\affiliation{Novosibirsk State University, Novosibirsk, Russia}
\author{M.~Turner}
\affiliation{CERN, Geneva, Switzerland}
\author{F.~Velotti}
\affiliation{CERN, Geneva, Switzerland}
\author{L.~Verra}
\affiliation{CERN, Geneva, Switzerland}
\affiliation{University of Milan, Milan, Italy}
\author{V.A.~Verzilov}
\affiliation{TRIUMF, Vancouver, Canada} 
\author{J.~Vieira}
\affiliation{GoLP/Instituto de Plasmas e Fus\~{a}o Nuclear, Instituto Superior T\'{e}cnico, Universidade de Lisboa, Lisbon, Portugal}
\author{C.P.~Welsch}
\affiliation{University of Liverpool, Liverpool, UK}
\affiliation{Cockcroft Institute, Daresbury, UK}
\author{M.~Wendt}
\affiliation{CERN, Geneva, Switzerland}
\author{B.~Williamson}
\affiliation{University of Manchester, Manchester, UK}
\affiliation{Cockcroft Institute, Daresbury, UK}
\author{M.~Wing}
\affiliation{UCL, London, UK}
\author{B.~Woolley}
\affiliation{CERN, Geneva, Switzerland}
\author{G.~Xia}
\affiliation{University of Manchester, Manchester, UK}
\affiliation{Cockcroft Institute, Daresbury, UK}
\author{G.~Zevi Della Porta}
\affiliation{CERN, Geneva, Switzerland}

\collaboration{The AWAKE Collaboration}
\noaffiliation


\date{\today}

\begin{abstract}
Plasma wakefield acceleration is a method for accelerating particle beams using electromagnetic fields that are orders of magnitude larger than those found in conventional radio frequency cavities. The core component of a plasma wakefield accelerator is the plasma source, which ranges from millimeter-scale gas jets used in laser-driven experiments, to the ten-meter-long rubidium cell used in the AWAKE experiment. The density of the neutral gas is a controlled input to the experiment, but the density of the plasma after ionization depends on many factors. AWAKE uses a high-energy proton beam to drive the plasma wakefield, and the wakefield acts back on the proton bunch by modulating it at the plasma frequency. We infer the plasma density by measuring the frequency of modulation of the proton bunch, and we measure the evolution of the density versus time by varying the arrival of the proton beam with respect to the ionizing laser pulse. Using this technique, we uncover a microsecond-long period of a stable plasma density followed by a rapid decay in density. The stability of the plasma after ionization has implications for the design of much longer vapor cells that could be used to accelerate particle beams to extremely high energies.

\end{abstract}

\pacs{Valid PACS appear here}

\maketitle

\section{\label{sec:intro}Introduction}

The Advanced Wakefield Experiment (AWAKE) at CERN is the first plasma wakefield acceleration (PWFA) experiment to use a high-energy proton beam driver~\cite{Assmann:2014hva,Caldwell:2015rkk,Gschwendtner:2015rni,Muggli:2017rkx}. The 400 GeV proton beam from the Super Proton Synchrotron (SPS) accelerator contains approximately $3\times 10^{11}$ particles per bunch, and the total stored energy in the beam is nearly 20 kJ, three orders of magnitude greater than the energy of electron beam drivers~\cite{Litos:2014} and laser pulse drivers~\cite{Gonsalves8GeV} used in state-of-art plasma acceleration experiments. The large amount of energy stored in the proton beam has the potential to simplify the design of a plasma wakefield accelerator: a trailing beam of electrons can be accelerated to extremely high energies in a single plasma stage, eliminating the need for complicated staging mechanisms~\cite{Caldwell:2008ak,Adli:2013npa,PhysRevSTAB.13.101301}. On the other hand, a proton beam-driven plasma accelerator must be hundreds of meters long in order to extract a significant fraction of the drive beam energy, which motivates the development of novel, scalable plasma sources with the capability of controlling and diagnosing the plasma density with sub-percent accuracy. While optical interferometry is commonly used to measure plasma densities in excess of $10^{17}$ cm$^{-3}$~\cite{DownerRMP}, no accurate diagnostics exist for the low density plasmas ($10^{14}-10^{15}$ cm$^{-3}$) used in the AWAKE experiment.


In typical PWFA experiments, the drive bunch is shorter than the plasma wavelength and can drive a high-amplitude wakefield as soon as it enters the plasma, but the proton bunch at AWAKE is too long to effectively couple to the plasma wave. However, the proton beam excites a self-modulation instability (SMI) as it passes through the plasma, which causes the beam to form micro-bunches separated by the plasma wavelength~\cite{Kumar:2010,Schroeder:2011,Pukhov:2011}. The micro-bunches act together to resonantly drive a high-amplitude wakefield. 

In order to control the onset of SMI, we use a high-power laser pulse to seed the instability. This process is referred to as seeded self-modulation (SSM)~\cite{Lotov:2015}. The seed pulse passes through and ionizes a rubidium (Rb) vapor source coincident with the passage of the proton beam. The seed pulse is temporally centered within the proton bunch, such that the head of the bunch transits through vapor and the tail transits through plasma. As the SSM develops, the micro-bunches and the plasma wave driven by the bunches have a fixed phase with respect to the seed pulse. The SSM phenomenon was recently demonstrated in a series of experiments at AWAKE~\cite{Turner:2018,Rieger:2018}.

The onset of micro-bunching depends on the local plasma density and the electric field strength of the proton beam. In the SSM process, both of these factors are controlled by the laser seed pulse, which ionizes the Rb at a specific co-moving longitudinal position, and therefore field strength, within the bunch. If plasma is present ahead of the seed pulse, it has the potential to disrupt the seeding process. It is therefore important to understand the degree to which a pre-ionized plasma affects the proton beam and how long the plasma persists after ionization, as this will limit the repetition rate of a plasma accelerator based on the SSM mechanism. Further, we are interested in the spatio-temporal properties of a long, laser-ionized plasma column, as this influences the design of future proton beam-driven plasma accelerators. 

In this work, we report on the lifetime and evolution of the plasma channel by inferring the plasma density from the frequency of a self-modulated proton bunch.


\section{\label{sec:experiment}Experimental Overview}

\begin{figure*}[ht]
	\centering\includegraphics[width=0.95\textwidth]{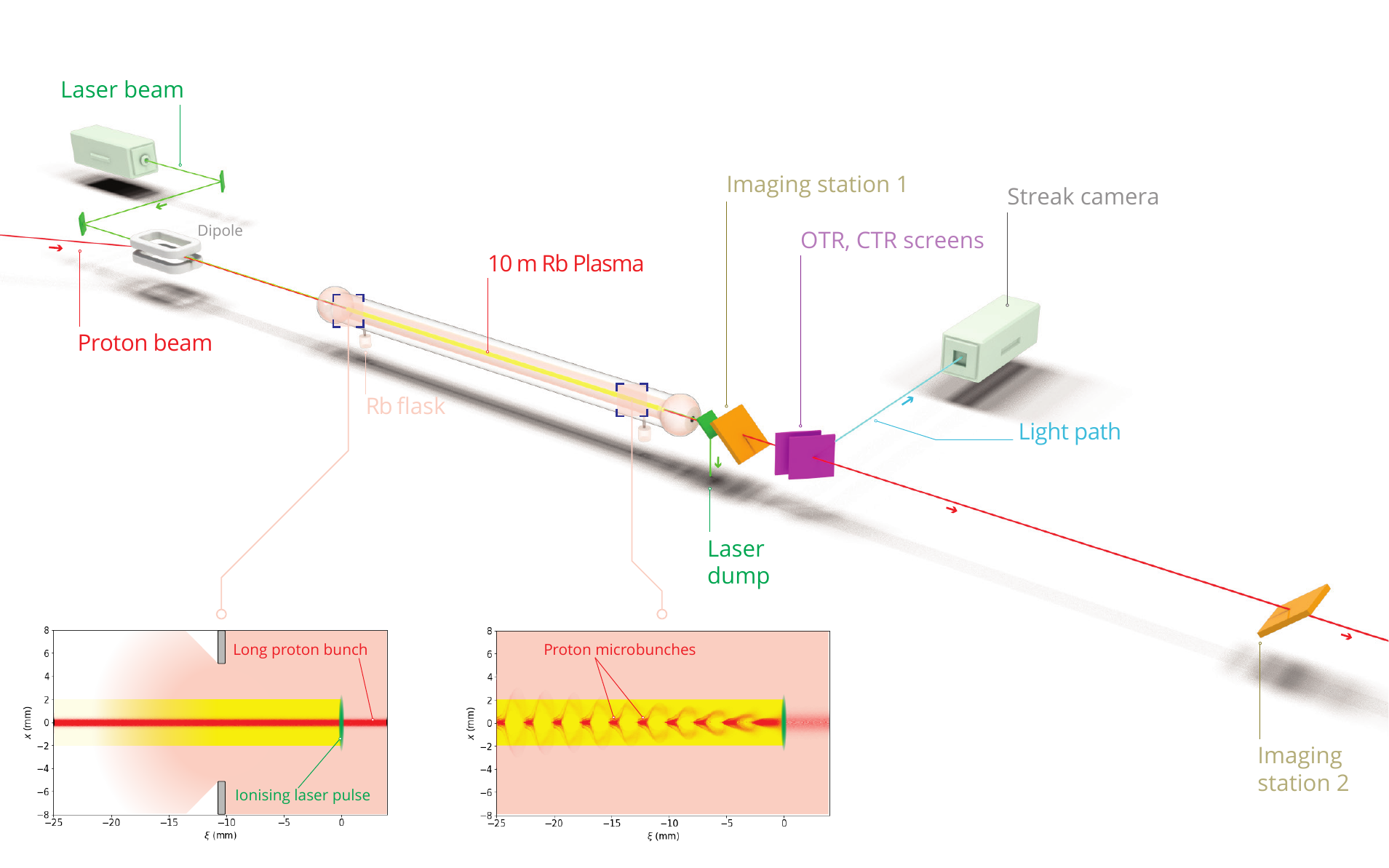}
	\caption{Experimental layout of the AWAKE experiment showing key components. Dipole magnets and optical mirrors are used to align the proton and laser beams upstream of the vapor cell. Downstream of the cell, the proton beam passes through an OTR foil and optical light is generated in the process. The light is transported to a streak camera for temporal imaging. The inset illustrates a uniform, Gaussian proton bunch entering the plasma which is micro-bunched by the time it exits the plasma.}
    \label{fig:layout}
\end{figure*}

The 400 GeV proton beam from the SPS is delivered to the AWAKE experiment via an 800~m beamline. For the dataset considered here, the proton beam contained $3.03 \pm 0.16 \times 10^{11}$ particles per bunch and a bunch length $\sigma_z = 8.24 \pm 0.15$ cm ($\sigma_t = 275 \pm 5$ ps). The transverse emittance was approximately 3.5 mm$\cdot$mrad in both planes. The beam is focused to a transverse spot size $\sigma_r \approx 200$ $\mu$m at the entrance of the vapor cell. A sketch of the experimental layout highlighting the key components is shown in Figure~\ref{fig:layout}.

The core of the AWAKE experiment is the 10 meter long Rb vapor cell~\cite{GPlyushc2018,Oz:2014}. Rb vapor flows into the cell from reservoirs at either end of the chamber. The heating of Rb reservoirs is adjusted until the vapor density is the same at both ends of the cell. The uniform heating of the cell implies that if the density is the same at both ends, it is uniform throughout. There are diagnostic ports at both ends of the cell where a white-light laser passes transversely through the vapor cell. The density of the vapor is inferred from spectroscopic measurements of white-light absorption at the 780 nm and 795 nm absorption lines, with an uncertainty of 0.5\%~\cite{Oz:2015agu,Batsch2018}.  The temperature of the cell was held at 468 K for this measurement.

The Rb vapor is ionized by a terawatt-class Ti:sapphire laser. The laser pulse energy can be varied from 40 to 450 mJ and the pulse length is approximately 120 fs~\cite{Fedosseev:2016ccm}. The laser is focused with a full-width half-maximum spot size of $2.05\pm0.05$ mm in $x$ and $1.20\pm0.02$ mm in $y$ near the entrance of the vapor cell. The laser-ionized plasma has a guiding effect on the laser pulse, which results in a small variation in spot size over the length of the cell. 
The main laser pulse is derived from an 88 MHz oscillator that is phase-locked to the RF cavities in the SPS ring~\cite{Damerau:IPAC2016-THPMY039}. This allows for synchronization of the amplified laser pulse with the extracted proton beam with picosecond accuracy.

The principal diagnostic for the measurements described in this paper is a streak camera. The proton bunch passes through a metallic foil two meters downstream of the exit of the Rb cell. The beam produces optical transition radiation (OTR) as it passes through the foil. The OTR light has the same spatio-temporal pattern as the micro-bunched beam. The light is sent via an optical transport system to a dark room where it is imaged onto the aperture of a streak camera. Inside the streak camera, the OTR photons are converted into electrons by a photocathode and accelerated through a ``streak tube'' towards a phosphor screen~\cite{hamamatsu}. A time-varying, transverse voltage is applied to the streak tube such that the electrons receive a kick that depends on when they were produced at the photocathode. The electrons arrive at different transverse positions on the phosphor screen and the emitted light is imaged onto a CMOS camera. The streak camera manufacturer provides a calibration of the streak axis so that the vertical dimension of the image is mapped to the co-moving temporal dimension of the beam. The AWAKE streak camera is a Hamamatsu C10910-05 model with a 16-bit, 2048$\times$2048 pixel ORCA-Flash4.0 CMOS sensor, which is binned $2\times2$ with a reduced region of interest (ROI) to produce a 508$\times$672 pixel image when operating in streak mode. The streak camera provides time windows ranging from several ns down to 73 ps. For the smallest time window, the resolution is limited by the streak dynamics rather than the pixel size and is 1 ps~\cite{Rieger:2017}. An optical target (Ronchi ruling with 5 lines per mm) was placed at the location of the foil and imaged by the streak camera to determine the point spread function (PSF) in the non-streaked (transverse) plane. The full-width half-maximum (FWHM) resolution in the transverse plane was determined to be 0.187 mm, or roughly 8 pixels.

\section{\label{sec:ana}Streak Image Analysis}

\begin{figure*}
   \centering
   \includegraphics*[width=\textwidth]{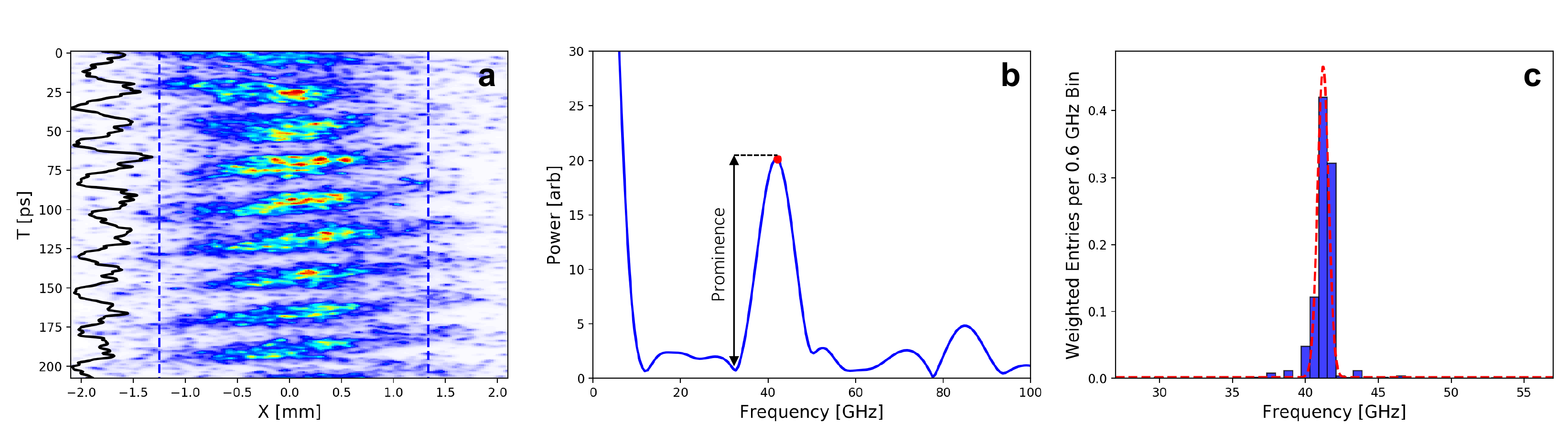}
   \caption{a) Streak camera image of a modulated proton beam in a 200 ps-long time window. The vertical axis is the time axis, with $t=0$ corresponding to the center of the proton bunch. The horizontal axis is the transverse spatial coordinate with $x=0$ located at the beam centroid. The ROI is denoted by dashed blue lines. The solid black line shows the temporal projection of the image data in the region of interest. b) The absolute value of the DFT of the projection shown in a) after application of the analysis procedure described in Section~\ref{sec:projection}. A peak in the spectrum is identified at $42.09\pm 0.6$ GHz using the peak-prominence algorithm and is denoted by the red dot, where the uncertainty of 0.6 GHz reflects the bin spacing of the frequency axis. The prominence is denoted by the black arrow. c) Histogram of the peak frequencies identified by the analysis described in Section~\ref{sec:column}. The entries to the histogram are weighted by their prominence and the histogram is fitted by a Gaussian centered at $41.57\pm0.07$ GHz, where the error on the centroid is extracted from the fit.}
   \label{fig:ana}
\end{figure*}

\subsection{\label{sec:projection} Fourier Transform of the Projection}
The streak camera was operated with a time window of 208 ps near the center of the proton bunch. The streak camera recorded images of the modulated bunch which were analyzed to extract the frequency of modulation. The first step of the analysis identifies the signal region in the streak camera image and takes a projection of the data along the time axis. The projection is the average of the observed charge density across each horizontal slice of the image. A Hann Filter is applied to the projection which is then embedded in a zero-padded array. A Discrete Fourier Transform (DFT) is taken of the zero-padded array. The purpose of zero-padding the array is to smooth the result of the DFT, while the Hann Filter prevents aliasing due to embedding. Figures~\ref{fig:ana}a) and~\ref{fig:ana}b) illustrate the signal identification and resulting DFT for a sample image. The modulation frequency is identified by a peak-prominence algorithm. The peak-prominence is defined as the amplitude of the local maximum with respect to the nearest local minimum on the low frequency side of the maximum. This definition reduces sensitivity to the high-amplitude DC component in the signal.

\subsection{\label{sec:column} Statistical Analysis}
The procedure described in Section~\ref{sec:projection} associates a single frequency with each streak camera image, but multiple frequencies are often observed across a single image. In the SSM phenomenon, focused protons appear at the center of the image and defocused protons appear at the edge of the image, phase-shifted by 180 degrees~\cite{Turner:2018}. The projection is an average of the on-axis and off-axis modulations which results in a reduced amplitude of the peak frequency.


In order to account for the variation of frequencies across an image, the image is subdivided into 40 columns, each 8 pixels wide, corresponding to the FWHM PSF resolution of the optical system. The DFT is taken for each column and the peak frequency is identified using the peak-prominence algorithm described above. The identified frequencies are added to a histogram, weighted by their prominence. Finally, a Gaussian is fitted to the resulting histogram, with the mean peak frequency and error on the mean extracted from the fit.

With the column-by-column approach, there is a risk of over-sampling the data and artificially reducing the statistical uncertainty. In the extreme case, where the columns are completely correlated with each other, using the column-by-column approach does not yield any additional information. To address this issue, we measure correlations between the frequencies in each column by generating a correlation matrix
\begin{equation}
    \mathrm{corr}_{ij} = \frac{1}{\sqrt{\sigma_i\sigma_j}} \sum_{n=1}^{n_{\mathrm{evt}}}\frac{w_{i,n}(f_{i,n}-\bar{f}_i)w_{j,n}(f_{j,n}-\bar{f}_j)}{\sum_n w_{i,n} \sum_n w_{j,n}},
\end{equation}
with $n_{\mathrm{evt}}$ the number of events in the sample, $w_{i,n}$ the weight (prominence) for column $i$ and event $n$, $f_{i,n}$ is the frequency identified for column $i$ and event $n$, $\bar{f}_i$ the average frequency in the column over $n_\mathrm{evt}$, and $\sigma_i$ the standard deviation of frequencies in the column over $n_\mathrm{evt}$. The correlation matrix is normalized such that the diagonal elements $\mathrm{corr}_{ii} = 1$. We assess the average correlation between adjacent columns by summing along the off-diagonals of the matrix. The first off-diagonal sum gives the average correlation between adjacent columns, the second off-diagonal sum gives the average between next-to-adjacent columns, etc. We compare correlation matrices calculated for single pixel-wide columns and 8 pixel-wide columns. In the single pixel-wide case, non-zero correlations are observed between adjacent and next-to-adjacent columns. In the 8 pixel-wide case, the correlation between adjacent columns is close to zero, which indicates that the frequency measurements in each column are independent.

\subsection{\label{sec:syst} Plasma Density Identification}
The measured microbunching frequency $f$ is determined by the action of the plasma wakefield on the proton bunch integrated over the length of the Rb vapor cell. The simplest interpretation of $f$ is that it corresponds to the plasma frequency $\omega_p/2\pi$, such that the plasma density $n$ is given by
\begin{equation} \label{eq:f2n}
n = \epsilon_0 m_e \left(\frac{2\pi f}{e}\right)^2,
\end{equation}
where $\epsilon_0$ is the permittivity of free space, $m_e$ is the mass of the electron, and $e$ is the charge of the electron. In previous work, we compared Rb vapor densities measured by the white-light interferometer with plasma densities derived from the microbunching frequency of the proton beam and found the two to be in agreement, which indicates a uniform, fully-ionized plasma~\cite{Rieger:2018}. 

There are a number of physical effects which may cause the measured microbunching frequency to differ from the underlying plasma frequency. These include longitudinal density gradients and frequency shifts due to the onset of microbunching. We are able to minimize the frequency shift due to the longitudinal density gradient through control of the vapor density. We estimate the frequency shift from this effect to be at most $\delta f/f = \delta n/2n \approx (9.5\pm1.0)\times 10^{-4}$~\cite{Petrenko2016}. The phase of the wakefield is affected by forced oscillations of the plasma at the onset of microbunching~\cite{Schroeder:2011,Pukhov:2011,Lotov:2015}. This effect is strongest just behind the seed point. In general, streak camera images are taken many periods behind the seed position where we expect the microbunching frequency to match the plasma frequency. The one exception is the dataset corresponding to zero time delay, which is described in more detail in Section~\ref{sec:Results}.





The standard error on the mean plasma density is given by
\begin{equation}
\sigma_{n} = 2n\frac{\sigma_{f}}{f},
\end{equation}
where $\sigma_{f}$ is the standard error on the mean microbunching frequency. A typical data sample is composed of 10 events, with 40 frequency measurements each, corresponding to the 40 columns of the streak camera image. The statistical uncertainty on the $\sim400$ measurements is typically $\sigma_{f} \approx 10^{-3}$, with a correspondingly small uncertainty on the plasma density. The statistical uncertainty on the plasma density achieved here is at least an order of magnitude better than what is achieved using optical probes of plasma density in PWFA experiments~\cite{DownerRMP}.

We can quantify the systematic uncertainty of the measurement in two ways. First, we compare the plasma density measurement at small time delays with the vapor density measurement, under the assumptions of complete ionization of the vapor and insufficient time for plasma recombination. For a time delay of 320 ps between a 135 mJ laser pulse and the center of the proton bunch, the measured plasma density is $n = (1.805\pm0.003)\times 10^{14}$ cm$^{-3}$, consistent with the measured vapor density is $n_{\mathrm{Rb}} = (1.810\pm0.009)\times 10^{14}$ cm$^{-3}$. 

For time delays of 500 ns or longer, the plasma has had time to relax and we can no longer compare densities derived from the microbunching frequency measurement to the Rb vapor density. To assess systematic errors at longer time delays, we compare different versions of the analysis. The average frequency can be calculated for each dataset (10 events grouped by time delay and laser pulse energy) by taking the projection of each streak image, computing the DFT of the projection, averaging the DFTs, and extracting the peak frequency. This approach produces one measurement per dataset. Alternatively, we can compute the DFT of the projection of each streak image, extract the peak frequency, and take the average over 10 measurements. Finally, we can use the column-by-column approach where the average is over 400 measurements. Comparing the single-measurement case with the column-by-column approach, we find that the difference in measured frequencies is 0.036 GHz averaged over all datasets. This is less than $\bar{\sigma}_{f400} = 0.167$ GHz, the average error of the column by column approach. Comparing the 10-measurement average with the column-by-column approach, we find that the difference in measured frequencies is 2.052 GHz averaged over all datasets, which is less than $\bar{\sigma}_{f10} = 3.567$ GHz, the average error of the projection method with 10 measurements per dataset. 

We conclude that the column-by-column analysis does not show a significant bias toward higher or lower plasma densities, when compared to other possible analyses or the Rb vapor density measurement. The measurement uncertainties quoted in the remainder of the paper reflect the statistical error on the mean from the column-by-column analysis.



\section{\label{sec:Results}Results}
\begin{figure*}[!tbh]
   \centering
   \includegraphics*[width=\textwidth]{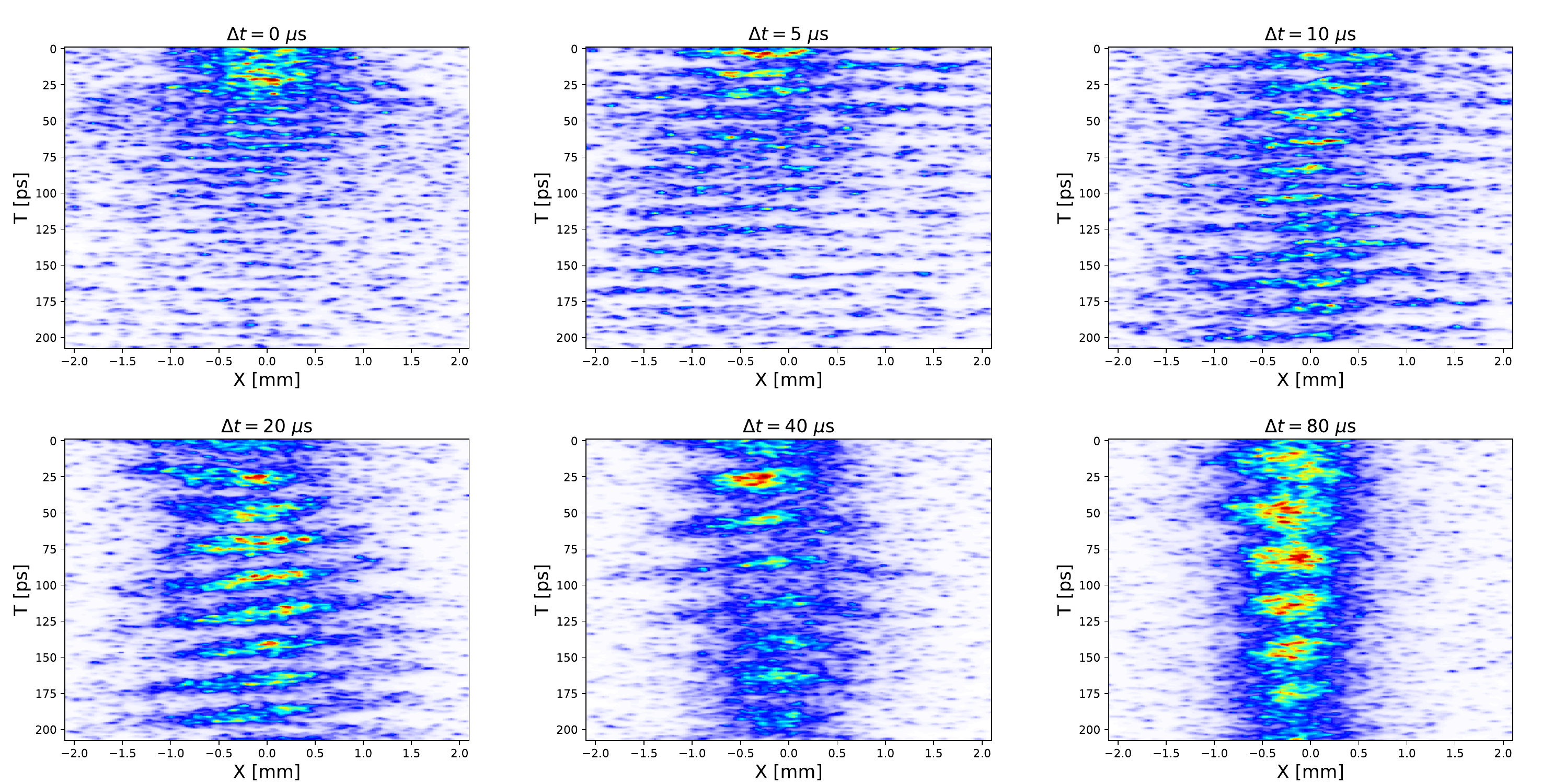}
   \caption{Streak camera images of the modulated proton beam for sample time delays $\Delta t$ after ionization by the laser pulse with the high-power setting. The frequency of the modulation is  observed to decrease with time after ionization, indicating a decay in the plasma density.}
   \label{fig:examples}
\end{figure*}

We study the evolution of the plasma density as a function of time after ionization at three laser pulse energies. The Rb vapor density is maintained at $(1.810 \pm 0.009) \times 10^{14}$ cm$^{-3}$, throughout the measurement. The relative laser--proton timing is adjusted via the phase-locking system mentioned in Section~\ref{sec:experiment}. The time delay $\Delta t$ is varied from 0 $\mu$s (corresponding to the laser seed pulse coincident with the center of the proton bunch) to 80 $\mu$s. The laser energy is controlled by changing the angle of an attenuating polarizer. For every time delay setting, we record ten shots each for a laser pulse energy of 135 mJ, 95 mJ, and 40 mJ, which we refer to as the high, medium, and low energy settings, respectively. The shot-to-shot energy variations and the transverse laser profile are measured on a ``virtual" laser line using leakage light from the mirror that folds the laser beam onto the proton beam axis. We record a single laser-off background shot for each setting, for an average of 33 events per time delay.  Figure~\ref{fig:examples} shows example streak camera images of the self-modulated proton bunches for different time delays after ionization with the high-power laser setting.

\begin{figure}[!tbh]
   \centering
   \includegraphics*[width=\columnwidth]{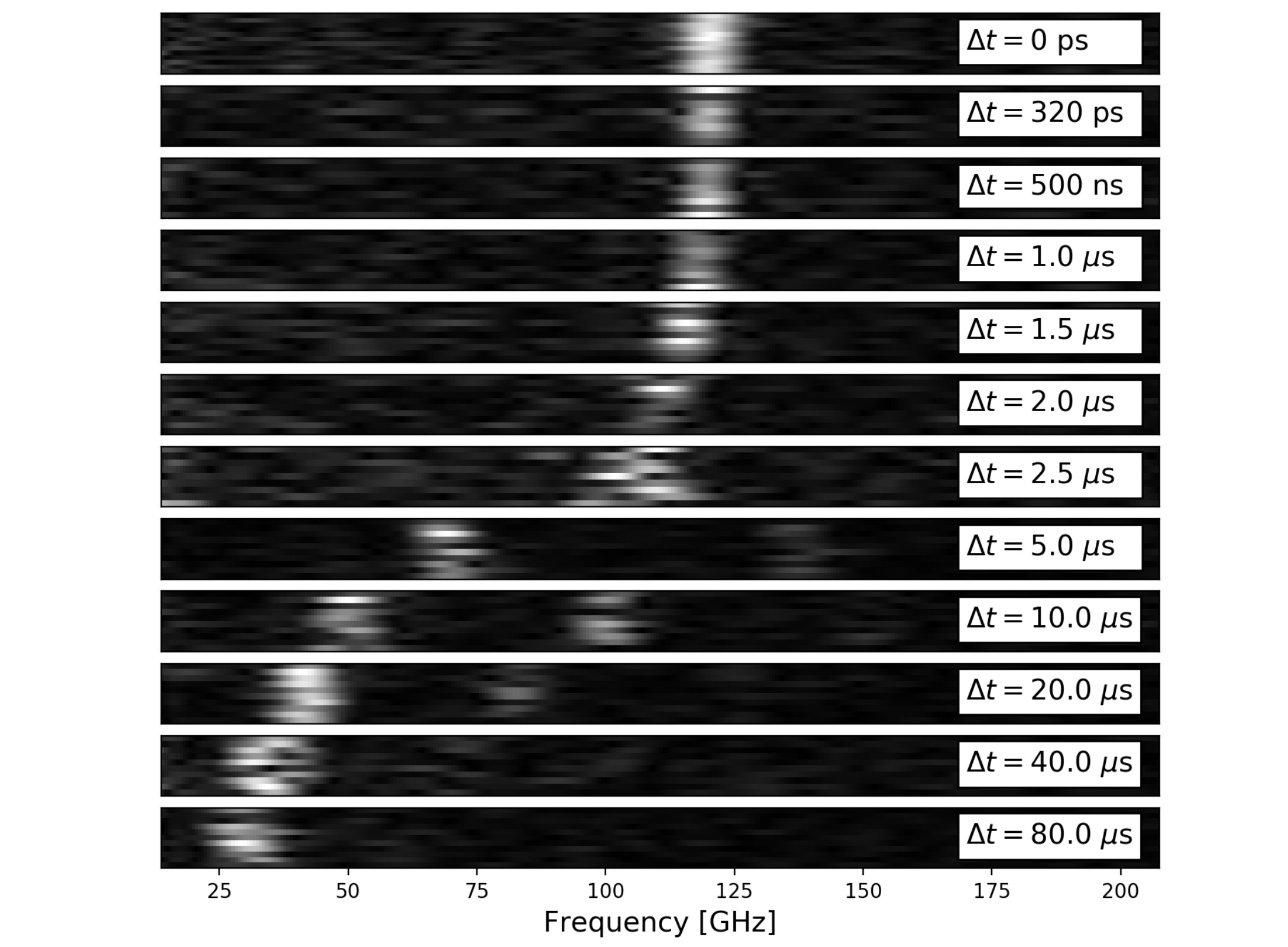}
   \caption{Fourier power spectra for the 135 mJ laser energy setting. Each subplot contains approximately 10 stacked Fourier spectra. The subplots are labeled by time after ionization, with 0 s at the top and 80 $\mu$s at the bottom. For time delays $\Delta t \geq 5$ $\mu$s, a second peak appears in the spectra at twice the frequency of the main peak.}
   \label{fig:waterfall}
\end{figure}

The streak camera images are grouped by time delay and laser energy setting. We take the DFT of the projection of each image (Section~\ref{sec:projection}) to estimate the peak frequency for each time delay and laser energy. The results of this analysis for the 135 mJ laser pulse energy setting are shown as a waterfall plot in Figure~\ref{fig:waterfall}. The data show a nearly constant modulation frequency over the first 1 $\mu$s, followed by rapid decay. Note that the larger time delays are logarithmically spaced. For time delays $\Delta t \geq 5~\mu$s, the second harmonic of the fundamental frequency is visible in the DFT spectrum. 

The peak frequencies and error on the frequencies are determined using the column-by-column analysis described in Section~\ref{sec:column}. The data are grouped into sets by time delay and laser pulse energy and the histograms for each image in the set are added together. The fit is performed on the ensemble histogram, with centroids and errors extracted from the fit.

\begin{figure}[!tbh]
   \centering
   \includegraphics*[width=\columnwidth]{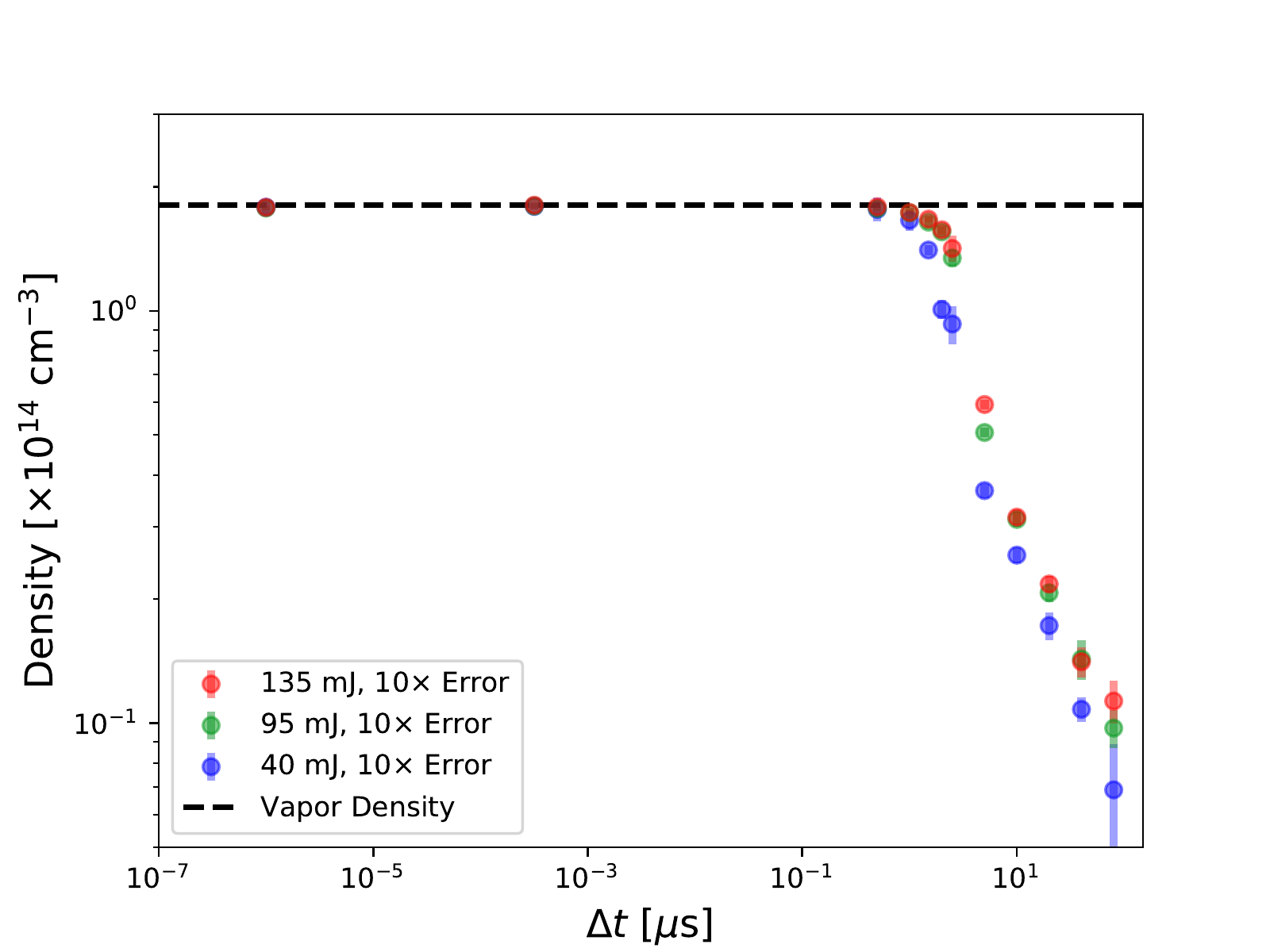}
   \caption{Plasma density versus time for the high, medium, and low energy laser settings, plotted on a log-log scale. The errors are shown multiplied by a factor of ten so that they are visible in the plot. The measured vapor density denoted by the black dashed line. The data points corresponding to $\Delta t = 0$ s are assigned an offset of 1 ps so that they are visible in the plot.}
   \label{fig:data}
\end{figure}

The results of the time delay scan are shown in Figure~\ref{fig:data}. The point corresponding to $\Delta t = 0$ s has been assigned an offset of 1 ps so that is visible on the log-scale plot and the uncertainties are multiplied by a factor of 10 for visibility. For a time delay of 0 s, corresponding to the laser co-propagating with the center of the proton bunch, the measured densities are $(1.783\pm0.002)\times 10^{14}$ cm$^{-3}$, $(1.777\pm0.002)\times 10^{14}$ cm$^{-3}$, and $(1.784\pm0.007)\times 10^{14}$ cm$^{-3}$ for the 135 mJ, 95 mJ, and 40 mJ laser pulse energy settings, respectively. These values are significantly less than the densities measured at a time delay of 320 ps ($(1.805\pm0.003)\times 10^{14}$ cm$^{-3}$, $(1.797\pm0.003)\times 10^{14}$ cm$^{-3}$, and $(1.793\pm0.003)\times 10^{14}$ cm$^{-3}$, for the 135 mJ, 95 mJ, and 40 mJ laser pulse energy settings, respectively), but we do not expect changes in the plasma density over such a short time scale. The lower density measured at $\Delta t = 0$ s may be attributed to a physical effect, the evolving phase velocity of the wake near the seed point discussed in Section~\ref{sec:syst}, and a measurement effect from the streak image which captures the proton bunch before and after the seed point. The unmodulated portion of the proton bunch can be seen towards the top of the $\Delta t = 0$ s image in Figure~\ref{fig:examples}. Neither of these effects are expected in the data with 320 ps delay. 


\section{\label{sec:model}Model and Analysis}


The data show three important features which we attempt to model. First, we observe a plateau in the plasma density lasting for up to a microsecond. The plateau is followed by an exponential decrease in the plasma density, and at long timescales, we observe a power-law like decay. Our model starts by estimating the initial conditions of the plasma immediately after formation by the ionizing laser pulse.

\subsection{Plasma Formation}\label{sec:ion}
The ionizing laser is focused into the Rb vapor cell with a FWHM spot size of 2.05 mm in $x$ and 1.20 mm in $y$. The peak intensity of the laser is $4.5\times 10^{13}$ W/cm$^2$, $3.2\times 10^{13}$ W/cm$^2$, $1.4\times 10^{13}$ W/cm$^2$, for the 135 mJ, 95 mJ, and 40 mJ laser energy settings, respectively. The ground state ionization energy of Rb is $U = 4.18$ eV, while the Ti:sapphire laser photon energy is $h\nu = 1.59$ eV at a central wavelength of 780 nm. In all cases, the Keldysh parameter is approximately 1
\begin{equation}
    \gamma_k = \frac{\omega\sqrt{2 m_e U}}{eE} \approx 1,
\end{equation}
where $\omega$ is the laser frequency and $E$ is the peak electric field~\cite{Keldysh1965}. For $\gamma_k \approx 1$, both multi-photon and field ionization mechanisms are present. We simulate the propagation of the ionizing laser pulse through the Rb vapor using a model that takes into account the nonlinear response of the vapor and multi-photon ionization~\cite{Gabor2019}. The simulated laser pulse parameters match that of the experiment, except that simulation is radially symmetric and the geometric mean of the FWHM in $x$ and $y$ is used as the radial size.

\begin{figure}[!tbh]
   \centering
   \includegraphics*[width=\columnwidth]{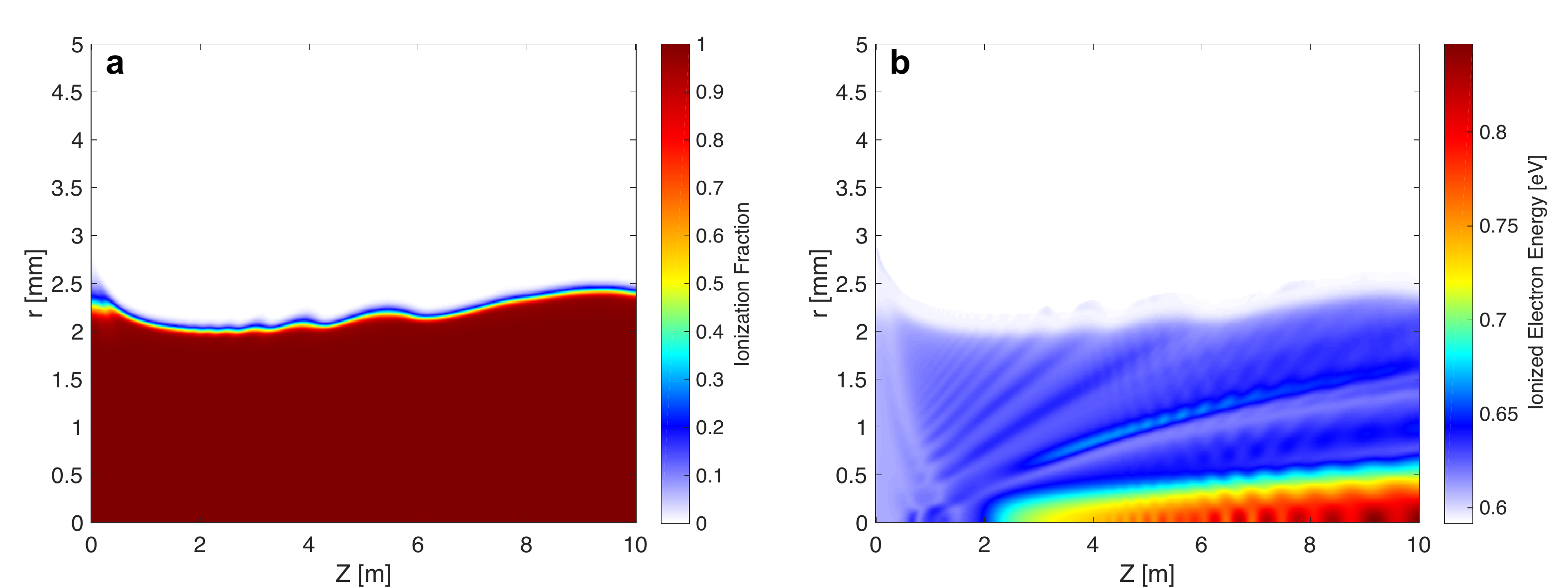}
   \caption{a) Ionization fraction for the 135 mJ laser pulse setting with a Gaussian transverse pulse shape based on the laser propagation model. The average radius of the ionized plasma column is 2.2 mm. b) Energy of electrons freed by ionization. The minimum energy after ionization is 0.59 eV, while the median energy is 0.62 eV.}
   \label{fig:ion}
\end{figure}

We are interested in the spatial distribution and temperature of the electrons freed through ionization, as this will serve as the initial conditions for our plasma evolution model. Figure~\ref{fig:ion}a) shows the radial and longitudinal ionization fraction of the Rb vapor along the 10 m length of the cell for the 135 mJ laser energy setting. The average radius of ionization is 2.2 mm. The transverse laser profile used in the model was Gaussian, but the laser mode profile in the experiment has a top-hat shape. Further simulations were performed with super-Gaussian modes. The results of these simulations show less uniformity at the boundary of the ionized region, but produce a similar value for the average ionized radius. Figure~\ref{fig:ion}b) shows the average energy of the plasma electrons in the ionized region. The minimum energy of the ionized electrons is 0.59 eV, corresponding to three-photon absorption $E_{min} = 3h\nu-U$. The median electron energy after ionization is 0.62 eV. The model does not include plasma electron energy gain in the ponderomotive potential of the laser pulse, which has a maximum amplitude of $U_p = 0.51$ eV. We expect the energy contribution from $U_p$ to be negligible given that the laser is linearly polarized and plasma electrons will be preferentially ionized at the maximum of the electric field which corresponds to the zero-crossing of the ponderomotive potential. Simulations were also performed for the 95 mJ and 40 mJ laser energy settings. At 95 mJ, the average ionized radius is 2.0 mm and at 40 mJ it is 1.7 mm. In all cases, the median energy of the plasma electrons after ionization is 0.62 eV.

\subsection{Thermalization, Diffusion, and Recombination}\label{sec:therm}
Several processes contribute to the evolution of the plasma column after ionization. Our model accounts for the following effects: 1) diffusion of electron thermal energy, 2) ambipolar plasma diffusion, 3) thermalization of electrons and ions, 4) thermalization of ions and neutral atoms, and 5) three-body plasma recombination. These processes are driven by collisions between electrons, ions and neutral particles. The fundamental parameter governing collisions in plasma is the electron-electron collision frequency due to Coulomb scattering~\cite{spitzer2006physics,fitzpatrick2015plasma,callen2006plasma}
\begin{equation}
    \nu_{ee} = \frac{4\sqrt{2\pi}}{3}\frac{e^4}{(4\pi\epsilon_0)^2 }\frac{n \ln{\Lambda}}{m_e^{1/2} T_e^{3/2}},
\end{equation}
where $\ln{\Lambda} = \ln{(12\pi n \lambda_d^3)}$ is the Coulomb logarithm and $\lambda_d$ the Debye length~\cite{FFChenIntro}. The inverse of the collision frequency, the collision time $\tau_{ee}$, gives the characteristic timescale for electrons to reach thermal equilibrium. For a density $n = 1.81\times 10^{14}$ cm$^{-3}$, and an initial electron energy $E = 0.62$ eV, $\tau_{ee} = 158$ ps. This is by far the fastest timescale in the model, and our model assumes that the electrons begin in thermal equilibrium with temperature $T_e = 2/3 E = 0.41$ eV.

The next relevant timescales are the ion-ion collision frequency $\nu_{ii} = \sqrt{m_e/m_i} \nu_{ee}$ and the electron-ion collision frequency $\nu_{ei} = m_e/m_i \nu_{ee}$. The mass of the Rb atom is 84.85 times the mass of the proton (155,000 times the mass of the electron), which allows us to neglect ion temperature when calculating collision frequencies. Ion-ion collisions are roughly 400 times slower than electron-electron collisions, and electron-ion collisions are roughly 155,000 times slower than electron-electron collisions.

The collision frequency between ions and neutrals is determined by the ion-neutral scattering cross section. The neutral atom cross section for elastic scattering is given by the size of the atom $\sigma_{Rb} = \pi r_{Rb}^2 \approx 2\times 10^{-15}$ cm$^2$. However, because the valence electron of the Rb atom is loosely bound, scattering is dominated by the charge-exchange cross section~\cite{RbCross1965}
\begin{equation}
    \sigma_{CEX} = (a-b\ln v_T)^2,
\end{equation}
where $a = 42\times 10^{-8}$ cm, $b = 1.85\times 10^{-8}$ cm and $v_T$ is the ion thermal velocity. For $v_T = 37,000$ cm/s, corresponding to an ion thermal energy of 0.041 eV, we have $\sigma_{CEX} = 5\times 10^{-14}$ cm$^2$, which is 25 times greater than the elastic scattering cross section. The ion-neutral collision frequency is given by $\nu_{i0} = n_n v_T \sigma_{CEX}$, where $n_n$ is the neutral atom density, and reaches a value of $3.42\times10^5$ s$^{-1}$ at the edge of the plasma column, which is comparable to the electron-ion collision frequency at the same location. Collisions between electrons and neutrals are neglected because they occur at the smaller cross section $\sigma_{Rb}$ and energy exchange between the two species is suppressed by a factor of $m_e/m_{Rb}$.

Three-body recombination occurs when two electrons collide in the vicinity of an ion, with one electron captured by the ion while the other carries away the excess momentum. The three-body recombination coefficient is given by~\cite{Gurevich1964}
\begin{equation}
    \alpha_3 = \frac{4\sqrt{2}\pi^{3/2}}{9}\frac{e^{10}}{(4\pi\epsilon_0)^5}\frac{\ln{\Lambda}}{m_e^{1/2}T_e^{9/2}} = 8.75\times10^{-27}(T_e [\mbox{eV}])^{-9/2}~\mbox{cm}^6/\mbox{s}.
\end{equation}
The three-body recombination rate has an extremely strong dependence on the electron temperature, and increases rapidly as the electrons cool. 

We are now in a position to describe the evolution of the plasma in terms of five quantities: the electron temperature $T_e(r,t)$, the ion temperature $T_i(r,t)$, the neutral temperature $T_0(r,t)$, the plasma density $n(r,t)$ and the neural density which is simply $n_0(r,t) = n_{init} - n(r,t)$. The evolution of the plasma density is governed by ambipolar diffusion and three-body recombination
\begin{equation}
    \frac{\del n}{\del t} = \nabla \cdot (D_a \nabla n) - \alpha_3 n^3, \label{eq:plas_dif}
\end{equation}
with the ambipolar diffusion constant $D_a = (1+T_e/T_i)D_i$, and $D_i = T_i/m_i \nu_{tot}$ and $\nu_{tot} = \nu_{ii} + \nu_{i0}$~\cite{FFChenIntro}. The ambipolar diffusion equation implies that the plasma is quasi-neutral so that $n(r,t)$ represents both the electron and ion densities. Inertial effects are estimated to be small and are not included in Equation~\ref{eq:plas_dif}.

The evolution of the electron temperature is given by
\begin{equation}
    \frac{3}{2}n\frac{\del T_e}{\del t} = \nabla \cdot (\kappa_e \nabla T_e) + 3\frac{m_e}{m_i} \nu_{ee} n (T_i-T_e), \label{eq:e_therm}
\end{equation}
where the electron thermal conductivity $\kappa_e = 3.2 n T_e/m_e\nu_{ee}$~\cite{fitzpatrick2015plasma}. The first term on the right hand side of the equation represents the thermal diffusion of electron temperature and the second term is the thermalization of electrons and ions.

Next, we have the ion temperature equation
\begin{equation}
\frac{3}{2}n\frac{\del T_i}{\del t} = \nabla \cdot (\kappa_i \nabla T_i) +3\frac{m_e}{m_i} \nu_{ee} n (T_e-T_i) + 2 \nu_{i0} n (T_0-T_i), \label{eq:ion_therm}
\end{equation}
where the ion thermal conductivity $\kappa_i = 3.9 n T_i/m_i\nu_{ii}$~\cite{fitzpatrick2015plasma}. Equation~\ref{eq:ion_therm} mirrors Equation~\ref{eq:e_therm}, with an additional term for ion-neutral thermalization. Finally, we include the heating of neutral atoms by ions
\begin{equation}
    \frac{\del T_0}{\del t} = 2 \nu_{i0} (T_i-T_0). \label{eq:neut_therm}
\end{equation}
The neutral temperature is also affected by the recombination process. When plasma ions and electrons combine, they add to the neutral density and increase the neutral temperature because the ions are typically warmer than the neutrals.

Equations~\ref{eq:plas_dif}-\ref{eq:neut_therm} are a system of coupled partial differential equations which must be solved numerically. The most straightforward approach to solving these equations is to use a forward-Euler method and choose short timesteps so that the Courant-Friedrichs-Lewy (CFL) condition is satisfied. The CFL condition requires $s = \Delta t \alpha/\Delta r^2 < 1/2$ at all grid points. However, this condition is nearly impossible to satisfy for Equation~\ref{eq:e_therm} given a reasonable time step and grid size. An alternative approach is to use a backward-Euler implicit method which is unconditionally stable and allows us to select a reasonable step size. In this case, a new complication arises because the coefficients of these equations ($D_a,~\kappa_e,~\nu_{ee}$) depend on $n$ and $T_e$ and must be recalculated at every step. We thus end up with a mismatch problem when the coefficients corresponding to time step $k$ are multiplying the dependent variable at time step $k+1$. We therefore use the implicit method but with a small time step to ensure that the coefficients vary slowly between iterations (\emph{e.g.} $D_a^{k}\approx D_a^{k+1})$.

The initial conditions are derived from the ionization model. The radial plasma density profile is computed by averaging the ionization fraction (Figure~\ref{fig:ion}a) along $z$, and the initial electron temperature is the average of the electron energy (Figure~\ref{fig:ion}b) along $z$, multiplied by a factor of two thirds. The boundary conditions for Equation~\ref{eq:plas_dif} are $\del n/\del r = 0 |_{r=0}$ and $\del n/\del r = 0 |_{r=r_w}$, where $r_w = 2$ cm is the radius of the pipe wall. We impose a minimum plasma density $n_{\mathrm{min}} = 10$ cm$^{-3}$ in order to avoid singularities in our system of equations. The value $n_{\mathrm{min}} = 10$ cm$^{-3}$ was chosen because it is consistent with the equilibrium ionization value given by the Saha equation for $T = 0.041$ eV. We found that the specific value of $n_{\mathrm{min}}$ has no effect on the results when varying this parameter over 8 orders of magnitude. 

The boundary conditions for Equations~\ref{eq:e_therm} and~\ref{eq:ion_therm} depend on the interaction of the plasma with the wall of the vapor cell. When the plasma reaches the wall, the electrons will be absorbed more rapidly than the ions, leading to the formation of an ion sheath~\cite{FFChenIntro}. The ions exchange energy with the wall, but the electrons are insulated by the sheath. The rate of thermalization between the ions and the wall depend on the heat capacity and thermal conductivity of the wall, and in turn these parameters vary with temperature~\cite{Broks2005}. In our experiment, the formation of the plasma sheath is not instantaneous. The plasma ions travel outward at the ambipolar velocity $v_a = (3 T_e/m_i)^{1/2}$, requiring at least 15 $\mu$s to reach the boundary and form the sheath. During this time, hot electrons reaching the wall will be absorbed and the electron temperature decreases. We do not model this process in detail, but rather take as a free parameter the electron temperature at the wall. A value of $T_e = 0.13$ eV provides the best fit to the data.





\begin{figure}[!tbh]
   \centering
   \includegraphics*[width=\columnwidth]{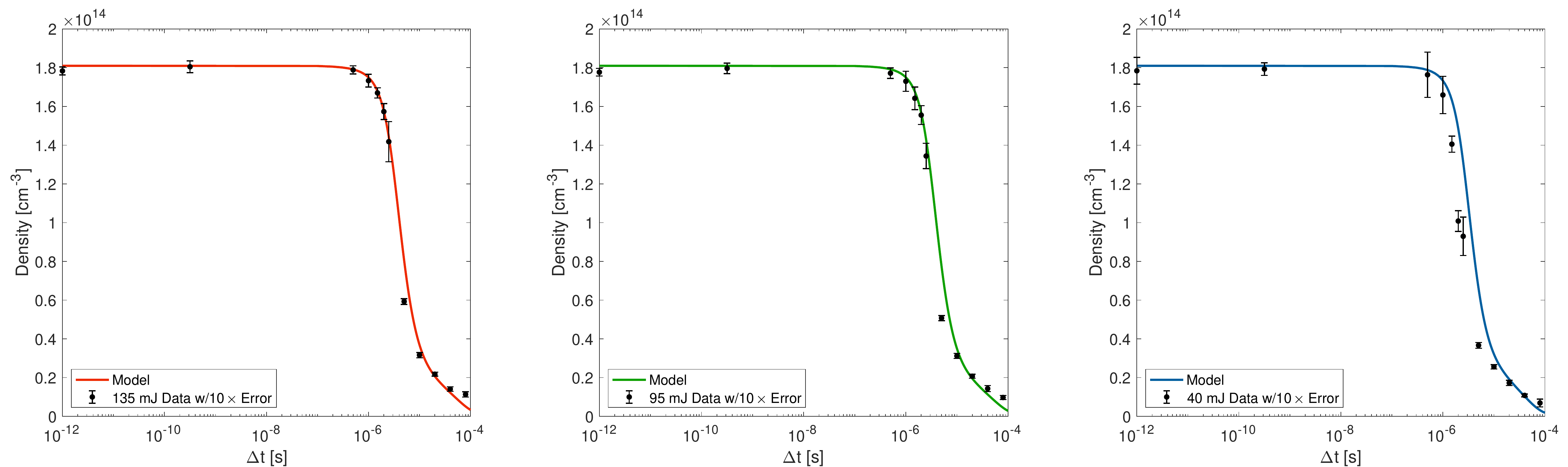}
   \caption{Plasma density versus log time compared to the on-axis density curves from the model. Error bars on the data are multiplied by a factor of 10 for visibility.}
   \label{fig:etch}
\end{figure}

The results of the model are shown in Figure~\ref{fig:etch}. There is excellent agreement between the model and the 135 mJ and 95 mJ data. The model overestimates the length of the plasma density plateau for the 40 mJ data, which indicates that the radius of the ionized region is less than 1.7 mm, the value predicted by the laser propagation model. A better fit can be achieved by assuming an initial radius of 1.1 mm. The discrepancy between the value predicted by the laser propagation model and what we observe in experiment is likely due to reduced confinement of the low-energy laser pulse. The high-energy laser pulse ionizes a wide channel with excellent self-confining properties, while the low-energy pulse ionizes a narrower channel which is more sensitive to asymmetries in the laser profile and leads to loss of confinement and further reduction of the channel width~\cite{Gabor2019}.

\section{Discussion and Summary}
In this work, we measure the decay of a singly-ionized Rubidium plasma as a function of time after ionization $0 < \Delta t < 80~\mu$s. The plasma density is inferred from the frequency of modulation of a high-energy proton beam. We observe three features in the data: a plateau in the plasma density for times $\Delta t \lesssim 1~\mu$s, a rapid decay in density for $1~\mu$s $< \Delta t < 5~\mu$s, and power-law decay thereafter. The data are well-characterized by a model which includes plasma diffusion, recombination, temperature diffusion, and intra-species thermalization.

This experiment was motivated by interest in the physics of the self-modulation instability, to understand optimal conditions for proton beam-driven plasma wakefield acceleration, and to gain understanding of the plasma relaxation rate which is useful for the design of future plasma sources for proton beam-driven plasma wakefield experiments. Of particular interest is the design of a ten meter long plasma source for AWAKE Run-II with uniform plasma density at the level of 0.25\%, which will be ionized by a back-propagating laser~\cite{2019arXiv191107534M}. The back-propagating laser implies a time delay of up to 66 ns between the time the laser ionizes the downstream end of the cell and the proton beam reaches the same location, assuming that the laser pulse and proton beam are coincident at the upstream end.  It is critical to understand whether or not the plasma density will evolve significantly during the 66 ns immediately after ionization by the laser. For time delays $\Delta t \ll \tau_{ei}$, the plasma electron temperature is roughly constant and therefore the three-body recombination rate is also constant. The fractional variation in the on-axis plasma density at short time delays is approximately
\begin{equation}
    \frac{\Delta n}{n} \approx \alpha_3 n^2 \Delta t.
\end{equation}
For a fixed density uniformity (\emph{e.g.} $\Delta n/n = 0.25$\%), the decay time scales as $n^{-2}$.

The AWAKE Run-II experiment is designed to operate at $7\times 10^{14}$ cm$^{-3}$ plasma density. Our model indicates that the plasma will decay by 0.25\% after 10 ns, which is shorter than the transit time of the laser pulse. However, the situation can be rectified by introducing a $1.7$\% gradient in the vapor density such that the downstream end, which is ionized first, has a higher initial plasma density than the upstream end. The gradient is chosen to compensate for plasma recombination such that the proton beam sees a uniform plasma density as it transits the cell. The radial variation in plasma density also meets the uniformity criterion up to $r = 1$ mm, which is much larger than the plasma skin depth and beam size $k_p^{-1} = \sigma_r = 200~\mu$m.

\acknowledgements
This work was supported in parts by the Russian Science Foundation (project 20-12-00062); a Leverhulme Trust Research Project Grant RPG- 2017-143; the STFC (AWAKE-UK, Cockroft Institute core and UCL consolidated grants), United Kingdom; a Deutsche Forschungsgemeinschaft project grant PU 213-6/1 “Three- dimensional quasi-static simulations of beam self-modulation for plasma wakefield acceleration”; the National Research Foundation of Korea (Nos. NRF-2015R1D1A1A01061074 and NRF- 2016R1A5A1013277); the Portuguese FCT — Foundation for Science and Technology, through grants CERN/FIS-TEC/0032/2017, PTDC-FIS-PLA-2940-2014, UID/FIS/50010/2013 and SFRH/IF/01635/2015; NSERC and CNRC for TRIUMF’s contribution; the EXMET program of the Hungarian Academy of Sciences, grant 2018-1.2.1-NKP-2018-00012 and the use of the MTA Cloud facility through the {\em Awakelaser} project; and the Research Council of Norway. M. Wing acknowledges the support of DESY, Hamburg. The AWAKE collaboration acknowledge the SPS team for their excellent proton delivery.

\bibliography{bibliography}

\begin{thebibliography}{35}%
\makeatletter
\providecommand \@ifxundefined [1]{%
 \@ifx{#1\undefined}
}%
\providecommand \@ifnum [1]{%
 \ifnum #1\expandafter \@firstoftwo
 \else \expandafter \@secondoftwo
 \fi
}%
\providecommand \@ifx [1]{%
 \ifx #1\expandafter \@firstoftwo
 \else \expandafter \@secondoftwo
 \fi
}%
\providecommand \natexlab [1]{#1}%
\providecommand \enquote  [1]{``#1''}%
\providecommand \bibnamefont  [1]{#1}%
\providecommand \bibfnamefont [1]{#1}%
\providecommand \citenamefont [1]{#1}%
\providecommand \href@noop [0]{\@secondoftwo}%
\providecommand \href [0]{\begingroup \@sanitize@url \@href}%
\providecommand \@href[1]{\@@startlink{#1}\@@href}%
\providecommand \@@href[1]{\endgroup#1\@@endlink}%
\providecommand \@sanitize@url [0]{\catcode `\\12\catcode `\$12\catcode
  `\&12\catcode `\#12\catcode `\^12\catcode `\_12\catcode `\%12\relax}%
\providecommand \@@startlink[1]{}%
\providecommand \@@endlink[0]{}%
\providecommand \url  [0]{\begingroup\@sanitize@url \@url }%
\providecommand \@url [1]{\endgroup\@href {#1}{\urlprefix }}%
\providecommand \urlprefix  [0]{URL }%
\providecommand \Eprint [0]{\href }%
\providecommand \doibase [0]{http://dx.doi.org/}%
\providecommand \selectlanguage [0]{\@gobble}%
\providecommand \bibinfo  [0]{\@secondoftwo}%
\providecommand \bibfield  [0]{\@secondoftwo}%
\providecommand \translation [1]{[#1]}%
\providecommand \BibitemOpen [0]{}%
\providecommand \bibitemStop [0]{}%
\providecommand \bibitemNoStop [0]{.\EOS\space}%
\providecommand \EOS [0]{\spacefactor3000\relax}%
\providecommand \BibitemShut  [1]{\csname bibitem#1\endcsname}%
\let\auto@bib@innerbib\@empty
\bibitem [{\citenamefont {Assmann}\ \emph {et~al.}(2014)\citenamefont {Assmann}
  \emph {et~al.}}]{Assmann:2014hva}%
  \BibitemOpen
  \bibfield  {author} {\bibinfo {author} {\bibfnamefont {R.}~\bibnamefont
  {Assmann}} \emph {et~al.} (\bibinfo {collaboration} {AWAKE Collaboration}),\
  }\href {\doibase 10.1088/0741-3335/56/8/084013} {\bibfield  {journal}
  {\bibinfo  {journal} {Plasma Phys. Control. Fusion}\ }\textbf {\bibinfo
  {volume} {56}},\ \bibinfo {pages} {084013} (\bibinfo {year}
  {2014})}\BibitemShut {NoStop}%
\bibitem [{\citenamefont {Caldwell}\ \emph {et~al.}(2016)\citenamefont
  {Caldwell} \emph {et~al.}}]{Caldwell:2015rkk}%
  \BibitemOpen
  \bibfield  {author} {\bibinfo {author} {\bibfnamefont {A.}~\bibnamefont
  {Caldwell}} \emph {et~al.} (\bibinfo {collaboration} {AWAKE Collaboration}),\
  }\href {\doibase 10.1016/j.nima.2015.12.050} {\bibfield  {journal} {\bibinfo
  {journal} {Nucl. Instrum. Meth.}\ }\textbf {\bibinfo {volume} {A829}},\
  \bibinfo {pages} {3} (\bibinfo {year} {2016})}\BibitemShut {NoStop}%
\bibitem [{\citenamefont {Gschwendtner}\ \emph {et~al.}(2016)\citenamefont
  {Gschwendtner} \emph {et~al.}}]{Gschwendtner:2015rni}%
  \BibitemOpen
  \bibfield  {author} {\bibinfo {author} {\bibfnamefont {E.}~\bibnamefont
  {Gschwendtner}} \emph {et~al.} (\bibinfo {collaboration} {AWAKE
  Collaboration}),\ }\href {\doibase
  https://doi.org/10.1016/j.nima.2016.02.026} {\bibfield  {journal} {\bibinfo
  {journal} {Nucl. Instrum. Meth.}\ }\textbf {\bibinfo {volume} {A829}},\
  \bibinfo {pages} {76} (\bibinfo {year} {2016})}\BibitemShut {NoStop}%
\bibitem [{\citenamefont {Muggli}\ \emph {et~al.}(2017)\citenamefont {Muggli}
  \emph {et~al.}}]{Muggli:2017rkx}%
  \BibitemOpen
  \bibfield  {author} {\bibinfo {author} {\bibfnamefont {P.}~\bibnamefont
  {Muggli}} \emph {et~al.} (\bibinfo {collaboration} {AWAKE Collaboration}),\
  }\href {\doibase 10.1088/1361-6587/aa941c} {\bibfield  {journal} {\bibinfo
  {journal} {Plasma Phys. Control. Fusion}\ }\textbf {\bibinfo {volume} {60}},\
  \bibinfo {pages} {014046} (\bibinfo {year} {2017})}\BibitemShut {NoStop}%
\bibitem [{\citenamefont {Litos}\ \emph {et~al.}(2014)\citenamefont {Litos}
  \emph {et~al.}}]{Litos:2014}%
  \BibitemOpen
  \bibfield  {author} {\bibinfo {author} {\bibfnamefont {M.}~\bibnamefont
  {Litos}} \emph {et~al.},\ }\href {\doibase 10.1038/nature13882} {\bibfield
  {journal} {\bibinfo  {journal} {Nature}\ }\textbf {\bibinfo {volume} {515}},\
  \bibinfo {pages} {92} (\bibinfo {year} {2014})}\BibitemShut {NoStop}%
\bibitem [{\citenamefont {Gonsalves}\ \emph {et~al.}(2019)\citenamefont
  {Gonsalves} \emph {et~al.}}]{Gonsalves8GeV}%
  \BibitemOpen
  \bibfield  {author} {\bibinfo {author} {\bibfnamefont {A.~J.}\ \bibnamefont
  {Gonsalves}} \emph {et~al.},\ }\href {\doibase
  10.1103/PhysRevLett.122.084801} {\bibfield  {journal} {\bibinfo  {journal}
  {Phys. Rev. Lett.}\ }\textbf {\bibinfo {volume} {122}},\ \bibinfo {pages}
  {084801} (\bibinfo {year} {2019})}\BibitemShut {NoStop}%
\bibitem [{\citenamefont {Caldwell}\ \emph {et~al.}(2009)\citenamefont
  {Caldwell} \emph {et~al.}}]{Caldwell:2008ak}%
  \BibitemOpen
  \bibfield  {author} {\bibinfo {author} {\bibfnamefont {A.}~\bibnamefont
  {Caldwell}} \emph {et~al.},\ }\href {\doibase 10.1038/NPHYS1248,
  10.1038/nphys1248} {\bibfield  {journal} {\bibinfo  {journal} {Nature Phys.}\
  }\textbf {\bibinfo {volume} {5}},\ \bibinfo {pages} {363} (\bibinfo {year}
  {2009})}\BibitemShut {NoStop}%
\bibitem [{\citenamefont {Adli}\ \emph {et~al.}(2013)\citenamefont {Adli} \emph
  {et~al.}}]{Adli:2013npa}%
  \BibitemOpen
  \bibfield  {author} {\bibinfo {author} {\bibfnamefont {E.}~\bibnamefont
  {Adli}} \emph {et~al.},\ }in\ \href
  {https://inspirehep.net/record/1246947/files/arXiv:1308.1145.pdf} {\emph
  {\bibinfo {booktitle} {{CSS: Minneapolis, USA, 2013}}}}\ (\bibinfo {year}
  {2013})\ \Eprint {http://arxiv.org/abs/1308.1145} {arXiv:1308.1145
  [physics.acc-ph]} \BibitemShut {NoStop}%
\bibitem [{\citenamefont {Schroeder}\ \emph {et~al.}(2010)\citenamefont
  {Schroeder} \emph {et~al.}}]{PhysRevSTAB.13.101301}%
  \BibitemOpen
  \bibfield  {author} {\bibinfo {author} {\bibfnamefont {C.~B.}\ \bibnamefont
  {Schroeder}} \emph {et~al.},\ }\href {\doibase 10.1103/PhysRevSTAB.13.101301}
  {\bibfield  {journal} {\bibinfo  {journal} {Phys. Rev. ST Accel. Beams}\
  }\textbf {\bibinfo {volume} {13}},\ \bibinfo {pages} {101301} (\bibinfo
  {year} {2010})}\BibitemShut {NoStop}%
\bibitem [{\citenamefont {Downer}\ \emph {et~al.}(2018)\citenamefont {Downer}
  \emph {et~al.}}]{DownerRMP}%
  \BibitemOpen
  \bibfield  {author} {\bibinfo {author} {\bibfnamefont {M.~C.}\ \bibnamefont
  {Downer}} \emph {et~al.},\ }\href {\doibase 10.1103/RevModPhys.90.035002}
  {\bibfield  {journal} {\bibinfo  {journal} {Rev. Mod. Phys.}\ }\textbf
  {\bibinfo {volume} {90}},\ \bibinfo {pages} {035002} (\bibinfo {year}
  {2018})}\BibitemShut {NoStop}%
\bibitem [{\citenamefont {Kumar}\ \emph {et~al.}(2010)\citenamefont {Kumar}
  \emph {et~al.}}]{Kumar:2010}%
  \BibitemOpen
  \bibfield  {author} {\bibinfo {author} {\bibfnamefont {N.}~\bibnamefont
  {Kumar}} \emph {et~al.},\ }\href {\doibase 10.1103/PhysRevLett.104.255003}
  {\bibfield  {journal} {\bibinfo  {journal} {Phys. Rev. Lett.}\ }\textbf
  {\bibinfo {volume} {104}},\ \bibinfo {pages} {255003} (\bibinfo {year}
  {2010})}\BibitemShut {NoStop}%
\bibitem [{\citenamefont {Schroeder}\ \emph {et~al.}(2011)\citenamefont
  {Schroeder} \emph {et~al.}}]{Schroeder:2011}%
  \BibitemOpen
  \bibfield  {author} {\bibinfo {author} {\bibfnamefont {C.~B.}\ \bibnamefont
  {Schroeder}} \emph {et~al.},\ }\href {\doibase
  10.1103/PhysRevLett.107.145002} {\bibfield  {journal} {\bibinfo  {journal}
  {Phys. Rev. Lett.}\ }\textbf {\bibinfo {volume} {107}},\ \bibinfo {pages}
  {145002} (\bibinfo {year} {2011})}\BibitemShut {NoStop}%
\bibitem [{\citenamefont {Pukhov}\ \emph {et~al.}(2011)\citenamefont {Pukhov}
  \emph {et~al.}}]{Pukhov:2011}%
  \BibitemOpen
  \bibfield  {author} {\bibinfo {author} {\bibfnamefont {A.}~\bibnamefont
  {Pukhov}} \emph {et~al.},\ }\href {\doibase 10.1103/PhysRevLett.107.145003}
  {\bibfield  {journal} {\bibinfo  {journal} {Phys. Rev. Lett.}\ }\textbf
  {\bibinfo {volume} {107}},\ \bibinfo {pages} {145003} (\bibinfo {year}
  {2011})}\BibitemShut {NoStop}%
\bibitem [{\citenamefont {Lotov}(2015)}]{Lotov:2015}%
  \BibitemOpen
  \bibfield  {author} {\bibinfo {author} {\bibfnamefont {K.~V.}\ \bibnamefont
  {Lotov}},\ }\href@noop {} {\bibfield  {journal} {\bibinfo  {journal} {Phys.
  Plasmas}\ }\textbf {\bibinfo {volume} {22}},\ \bibinfo {pages} {103110}
  (\bibinfo {year} {2015})}\BibitemShut {NoStop}%
\bibitem [{\citenamefont {Turner}\ \emph {et~al.}(2019)\citenamefont {Turner}
  \emph {et~al.}}]{Turner:2018}%
  \BibitemOpen
  \bibfield  {author} {\bibinfo {author} {\bibfnamefont {M.}~\bibnamefont
  {Turner}} \emph {et~al.} (\bibinfo {collaboration} {AWAKE Collaboration}),\
  }\href {\doibase 10.1103/PhysRevLett.122.054801} {\bibfield  {journal}
  {\bibinfo  {journal} {Phys. Rev. Lett.}\ }\textbf {\bibinfo {volume} {122}},\
  \bibinfo {pages} {054801} (\bibinfo {year} {2019})}\BibitemShut {NoStop}%
\bibitem [{\citenamefont {Adli}\ \emph {et~al.}(2019)\citenamefont {Adli} \emph
  {et~al.}}]{Rieger:2018}%
  \BibitemOpen
  \bibfield  {author} {\bibinfo {author} {\bibfnamefont {E.}~\bibnamefont
  {Adli}} \emph {et~al.} (\bibinfo {collaboration} {AWAKE Collaboration}),\
  }\href {\doibase 10.1103/PhysRevLett.122.054802} {\bibfield  {journal}
  {\bibinfo  {journal} {Phys. Rev. Lett.}\ }\textbf {\bibinfo {volume} {122}},\
  \bibinfo {pages} {054802} (\bibinfo {year} {2019})}\BibitemShut {NoStop}%
\bibitem [{\citenamefont {Plyushchev}\ \emph {et~al.}(2018)\citenamefont
  {Plyushchev} \emph {et~al.}}]{GPlyushc2018}%
  \BibitemOpen
  \bibfield  {author} {\bibinfo {author} {\bibfnamefont {G.}~\bibnamefont
  {Plyushchev}} \emph {et~al.},\ }\href
  {http://iopscience.iop.org/article/10.1088/1361-6463/aa9dd7/pdf} {\bibfield
  {journal} {\bibinfo  {journal} {J. Phys. D: Appl. Phys}\ }\textbf {\bibinfo
  {volume} {51}} (\bibinfo {year} {2018})}\BibitemShut {NoStop}%
\bibitem [{\citenamefont {{\"Oz}}\ \emph {et~al.}(2014)\citenamefont {{\"Oz}}
  \emph {et~al.}}]{Oz:2014}%
  \BibitemOpen
  \bibfield  {author} {\bibinfo {author} {\bibfnamefont {E.}~\bibnamefont
  {{\"Oz}}} \emph {et~al.},\ }\href {\doibase
  https://doi.org/10.1016/j.nima.2013.10.093} {\bibfield  {journal} {\bibinfo
  {journal} {Nucl. Instrum. Meth.}\ }\textbf {\bibinfo {volume} {A740}},\
  \bibinfo {pages} {197} (\bibinfo {year} {2014})}\BibitemShut {NoStop}%
\bibitem [{\citenamefont {{\"Oz}}\ \emph {et~al.}(2016)\citenamefont {{\"Oz}}
  \emph {et~al.}}]{Oz:2015agu}%
  \BibitemOpen
  \bibfield  {author} {\bibinfo {author} {\bibfnamefont {E.}~\bibnamefont
  {{\"Oz}}} \emph {et~al.},\ }\href {\doibase 10.1016/j.nima.2016.02.005}
  {\bibfield  {journal} {\bibinfo  {journal} {Nucl. Instrum. Meth.}\ }\textbf
  {\bibinfo {volume} {A829}},\ \bibinfo {pages} {321} (\bibinfo {year}
  {2016})}\BibitemShut {NoStop}%
\bibitem [{\citenamefont {Batsch}\ \emph {et~al.}(2018)\citenamefont {Batsch}
  \emph {et~al.}}]{Batsch2018}%
  \BibitemOpen
  \bibfield  {author} {\bibinfo {author} {\bibfnamefont {F.}~\bibnamefont
  {Batsch}} \emph {et~al.},\ }\href {\doibase 10.1016/j.nima.2018.02.067}
  {\bibfield  {journal} {\bibinfo  {journal} {Nucl. Instrum. Meth.}\ }\textbf
  {\bibinfo {volume} {909}},\ \bibinfo {pages} {359} (\bibinfo {year}
  {2018})}\BibitemShut {NoStop}%
\bibitem [{\citenamefont {Fedosseev}\ \emph {et~al.}()\citenamefont {Fedosseev}
  \emph {et~al.}}]{Fedosseev:2016ccm}%
  \BibitemOpen
  \bibfield  {author} {\bibinfo {author} {\bibfnamefont {V.}~\bibnamefont
  {Fedosseev}} \emph {et~al.},\ }in\ \href {\doibase
  10.18429/JACoW-IPAC2016-WEPMY020} {\emph {\bibinfo {booktitle} {{IPAC
  Proceedings: Busan, Korea, May 2016}}}},\ pp.\ \bibinfo {pages} {2592 --
  2595}\BibitemShut {NoStop}%
\bibitem [{\citenamefont {Damerau}\ \emph {et~al.}(2016)\citenamefont {Damerau}
  \emph {et~al.}}]{Damerau:IPAC2016-THPMY039}%
  \BibitemOpen
  \bibfield  {author} {\bibinfo {author} {\bibfnamefont {H.}~\bibnamefont
  {Damerau}} \emph {et~al.},\ }\href {\doibase
  10.18429/JACOW-IPAC2016-THPMY039} {\bibfield  {journal} {\bibinfo  {journal}
  {Proceedings of the 7th Int. Particle Accelerator Conf.}\ }\textbf {\bibinfo
  {volume} {IPAC2016}},\ \bibinfo {pages} {Korea} (\bibinfo {year}
  {2016})}\BibitemShut {NoStop}%
\bibitem [{\citenamefont {Hamamatsu}(2008)}]{hamamatsu}%
  \BibitemOpen
  \bibfield  {author} {\bibinfo {author} {\bibnamefont {Hamamatsu}},\ }\href
  {https://www.hamamatsu.com/resources/pdf/sys/SHSS0006E_STREAK.pdf} {\emph
  {\bibinfo {title} {Guide to Streak Cameras}}}\ (\bibinfo  {publisher}
  {Hamamatsu Corporation},\ \bibinfo {year} {2008})\BibitemShut {NoStop}%
\bibitem [{\citenamefont {Rieger}\ \emph {et~al.}(2017)\citenamefont {Rieger}
  \emph {et~al.}}]{Rieger:2017}%
  \BibitemOpen
  \bibfield  {author} {\bibinfo {author} {\bibfnamefont {K.}~\bibnamefont
  {Rieger}} \emph {et~al.},\ }\href@noop {} {\bibfield  {journal} {\bibinfo
  {journal} {Rev. Sci. Inst.}\ }\textbf {\bibinfo {volume} {88}},\ \bibinfo
  {pages} {025110} (\bibinfo {year} {2017})}\BibitemShut {NoStop}%
\bibitem [{\citenamefont {Petrenko}\ \emph {et~al.}(2016)\citenamefont
  {Petrenko} \emph {et~al.}}]{Petrenko2016}%
  \BibitemOpen
  \bibfield  {author} {\bibinfo {author} {\bibfnamefont {A.}~\bibnamefont
  {Petrenko}} \emph {et~al.},\ }\href {\doibase 10.1016/j.nima.2016.01.063}
  {\bibfield  {journal} {\bibinfo  {journal} {Nucl. Instrum Meth.}\ }\textbf
  {\bibinfo {volume} {829}},\ \bibinfo {pages} {63} (\bibinfo {year}
  {2016})}\BibitemShut {NoStop}%
\bibitem [{\citenamefont {Keldysh}(1965)}]{Keldysh1965}%
  \BibitemOpen
  \bibfield  {author} {\bibinfo {author} {\bibfnamefont {L.~V.}\ \bibnamefont
  {Keldysh}},\ }\href {http://www.jetp.ac.ru/cgi-bin/dn/e_020_05_1307.pdf}
  {\bibfield  {journal} {\bibinfo  {journal} {Sov. Phys. JETP}\ }\textbf
  {\bibinfo {volume} {20}},\ \bibinfo {pages} {1307} (\bibinfo {year}
  {1965})}\BibitemShut {NoStop}%
\bibitem [{\citenamefont {Demeter}(2019)}]{Gabor2019}%
  \BibitemOpen
  \bibfield  {author} {\bibinfo {author} {\bibfnamefont {G.}~\bibnamefont
  {Demeter}},\ }\href {\doibase 10.1103/PhysRevA.99.063423} {\bibfield
  {journal} {\bibinfo  {journal} {Phys. Rev. A}\ }\textbf {\bibinfo {volume}
  {99}},\ \bibinfo {pages} {063423} (\bibinfo {year} {2019})}\BibitemShut
  {NoStop}%
\bibitem [{\citenamefont {Spitzer}(2006)}]{spitzer2006physics}%
  \BibitemOpen
  \bibfield  {author} {\bibinfo {author} {\bibfnamefont {L.}~\bibnamefont
  {Spitzer}},\ }\href@noop {} {\emph {\bibinfo {title} {Physics of fully
  ionized gases}}}\ (\bibinfo  {publisher} {Dover Publications},\ \bibinfo
  {address} {Mineola, N.Y},\ \bibinfo {year} {2006})\BibitemShut {NoStop}%
\bibitem [{\citenamefont {Fitzpatrick}(2015)}]{fitzpatrick2015plasma}%
  \BibitemOpen
  \bibfield  {author} {\bibinfo {author} {\bibfnamefont {R.}~\bibnamefont
  {Fitzpatrick}},\ }\href@noop {} {\emph {\bibinfo {title} {Plasma physics : an
  introduction}}}\ (\bibinfo  {publisher} {CRC Press, Taylor \& Francis
  Group},\ \bibinfo {address} {Boca Raton, FL},\ \bibinfo {year}
  {2015})\BibitemShut {NoStop}%
\bibitem [{\citenamefont {Callen}(2006)}]{callen2006plasma}%
  \BibitemOpen
  \bibfield  {author} {\bibinfo {author} {\bibfnamefont {J.~D.}\ \bibnamefont
  {Callen}},\ }\href {http://homepages.cae.wisc.edu/~callen/book.html} {\emph
  {\bibinfo {title} {Fundamentals of Plasma Physics}}}\ (\bibinfo  {publisher}
  {University of Wisconsin},\ \bibinfo {address} {Madison, WI},\ \bibinfo
  {year} {2006})\BibitemShut {NoStop}%
\bibitem [{\citenamefont {Chen}(2016)}]{FFChenIntro}%
  \BibitemOpen
  \bibfield  {author} {\bibinfo {author} {\bibfnamefont {F.~F.}\ \bibnamefont
  {Chen}},\ }\href {\doibase 10.1007/978-3-319-22309-4} {\emph {\bibinfo
  {title} {Introduction to Plasma Physics and Controlled Fusion}}}\ (\bibinfo
  {publisher} {Springer},\ \bibinfo {year} {2016})\BibitemShut {NoStop}%
\bibitem [{\citenamefont {Perel}\ \emph {et~al.}(1965)\citenamefont {Perel}
  \emph {et~al.}}]{RbCross1965}%
  \BibitemOpen
  \bibfield  {author} {\bibinfo {author} {\bibfnamefont {J.}~\bibnamefont
  {Perel}} \emph {et~al.},\ }\href {\doibase 10.1103/PhysRev.138.A937}
  {\bibfield  {journal} {\bibinfo  {journal} {Phys. Rev.}\ }\textbf {\bibinfo
  {volume} {138}},\ \bibinfo {pages} {A937} (\bibinfo {year}
  {1965})}\BibitemShut {NoStop}%
\bibitem [{\citenamefont {Gurevich}\ and\ \citenamefont
  {Pitaevskii}(1964)}]{Gurevich1964}%
  \BibitemOpen
  \bibfield  {author} {\bibinfo {author} {\bibfnamefont {A.~V.}\ \bibnamefont
  {Gurevich}}\ and\ \bibinfo {author} {\bibfnamefont {L.~P.}\ \bibnamefont
  {Pitaevskii}},\ }\href {http://jetp.ac.ru/cgi-bin/e/index/e/19/4/p870}
  {\bibfield  {journal} {\bibinfo  {journal} {Soviet Physics JETP}\ }\textbf
  {\bibinfo {volume} {19}},\ \bibinfo {pages} {870} (\bibinfo {year}
  {1964})}\BibitemShut {NoStop}%
\bibitem [{\citenamefont {Broks}\ \emph {et~al.}(2005)\citenamefont {Broks}
  \emph {et~al.}}]{Broks2005}%
  \BibitemOpen
  \bibfield  {author} {\bibinfo {author} {\bibfnamefont {B.~H.~P.}\
  \bibnamefont {Broks}} \emph {et~al.},\ }\href {\doibase
  10.1103/physreve.71.016401} {\bibfield  {journal} {\bibinfo  {journal}
  {Physical Review E}\ }\textbf {\bibinfo {volume} {71}} (\bibinfo {year}
  {2005}),\ 10.1103/physreve.71.016401}\BibitemShut {NoStop}%
\bibitem [{\citenamefont {{Muggli}}(2019)}]{2019arXiv191107534M}%
  \BibitemOpen
  \bibfield  {author} {\bibinfo {author} {\bibfnamefont {P.}~\bibnamefont
  {{Muggli}}},\ }\href@noop {} {\enquote {\bibinfo {title} {{Physics to plan
  AWAKE Run 2}},}\ } (\bibinfo {year} {2019}),\ \Eprint
  {http://arxiv.org/abs/1911.07534} {arXiv:1911.07534 [physics.acc-ph]}
  \BibitemShut {NoStop}%
\end{thebibliography}%

\end{document}